\documentclass{elsart}
\usepackage{graphics}
\usepackage{graphicx}
\usepackage{epsfig}
\usepackage{amssymb,amsmath}
\usepackage[colorlinks]{hyperref}
\usepackage{rotating}
\usepackage{tikz}
\newtheorem{definition}{Definition}[section]

\begin{document}
\linespread{1}
\begin{frontmatter}
%\bibliographystyle{fr-plain}
\title{Towards New Multiwavelets: Associated Filters and Algorithms. Part I: Theoretical Framework and Investigation of Biomedical Signals, ECG and Coronavirus Cases}
\author{Malika Jallouli}
\address{Universit\'e de Sousse, Ecole Nationale d'Ingénieurs de Sousse, LATIS Laboratory of Advanced Technology and Intelligent Systems, 4023 Sousse, Tunisie.} \ead{jallouli.malika3@gmail.com}
\author{Makerem Zemni}
\address{Universit\'e de Sousse, Ecole Nationale d'Ingénieurs de Sousse, LATIS Laboratory of Advanced Technology and Intelligent Systems, 4023 Sousse, Tunisie.} \ead{makaremzemni@gmail.com}
\author{Anouar Ben Mabrouk\corauthref{Label1}}
\address{Department of Mathematics, Higher Institute of Applied Mathematics and Computer Science, Street of Assad Ibn Alfourat, 3100 Kairouan, Tunisia.\\
\& Laboratory of Algebra, Number Theory and Nonlinear Analysis, LR18ES15, Department of Mathematics, Faculty of Sciences, 5000 Monastir, Tunisia.\\
\& Department of Mathematics, Faculty of Sciences, University of Tabuk, Saudi Arabia.}
\ead{anouar.benmabrouk@fsm.rnu.tn}
\corauth[Label1]{Corresponding author.}
\author{Mohamed Ali Mahjoub}
\address{Universit\'e de Sousse, Ecole Nationale d'Ing\'enieurs de Sousse, LATIS - Laboratory of Advanced Technology and Intelligent Systems, 4023, Sousse, Tunisie.}	
\ead{medalimahjoub@gmail.com}           %  \\
%
%%%%%%%%%%%%%%%%%%%%%%%%%%%%%%%%%%%%%%%%%%%%%%%%%%%%%%%%%%%%%%%%%%%%%%%%%%%%%%%%%%%%%%%%%%%%%%%%%%%%%%%%%%%%%%%%%%%%%%%%%%%%%%%%%%%%%%%%%%%%%%%%
%%%%%%%%%%%%%%%%%%%%%%%%%%%%%%%%%%%%%%%%%%%%%%%%%%%%%%%%%%%%%%%%%%%%%%%%%%%%%%%%%%%%%%%%%%%%%%%%%%%%%%%%%%%%%%%%%%%%%%%%%%%%%%%%%%%%%%%%%%%%%%%%
%%%%%%%%%%%%%%%%%%%%%%%%%%%%%%%%%%%%%%%%%%%%%%%%%%%%%%%%%%%%%%%%%%%%%%%%%%%%%%%%%%%%%%%%%%%%%%%%%%%%%%%%%%%%%%%%%%%%%%%%%%%%%%%%%%%%%%%%%%%%%%%%
%%%%%%%%%%%%%%%%%%%%%%%%%%%%%%%%%%%%%%%%%%%%%%%%%%%%%%%%%%%%%%%%%%%%%%%%%%%%%%%%%%%%%%%%%%%%%%%%%%%%%%%%%%%%%%%%%%%%%%%%%%%%%%%%%%%%%%%%%%%%%%%%
\begin{abstract}
Biosignals are nowadays important subjects for scientific researches from both theory and applications especially with the appearance of new pandemics threatening the humanity such as the new Coronavirus. One aim in the present work is to prove that Wavelets may be a successful machinery to understand such phenomena by applying a step forward extension of wavelets to multiwavelets. We proposed in a first step to improve multiwavelet notion by constructing more general families using independent components for multi-scaling and multiwavelet mother functions. A special multiwavelet is then introduced, continuous and discrete multiwavelet transforms are associated, as well as new filters and algorithms of decomposition and reconstruction. The constructed multiwavelet framework is applied for some experimentations showing fast algorithms, ECG signal and a strain of Coronavirus processing.
\end{abstract}
\end{frontmatter}
%\tableofcontents
%
\section{Introduction}
Image processing is a discipline of computer science and applied mathematics that studies digital images and their transformations, with the aim of improving their quality or extracting information from them and in applied fields such as medicine, geography, town planning to model and understand real life.

This is a subset of signal processing dedicated to images and derived data such as video (as opposed to parts of signal processing devoted to other types of data: notably sound and other one-dimensional signals), while operating in the digital domain (as opposed to analog signal processing techniques, such as traditional photography or television). The image processing must be carried out purely digital, leading to a well description, sophisticated analysis and ease application in quantitative forms.

In theory, image/signal processing is a domain of science that is not recent, but in contrast, it is developed till the early discovery of Fourier analysis. Two large concepts are distinguished. Classical image/signal processing known as analogical and digital or numerical image/signal processing which is the current revolution in the field. It consists of a box of techniques and/or methods, mathematical and/or physical, theoretical and/or applied that aim to modify or to convert an image in another form in order to improve it and/or to extract information. As a result, the image processing is the set of methods and techniques operating on them, in order to make operations possible, easy, more effective, to improve the visual appearance of the image and to extract relevant informations (See \cite{Mallat,Kotas,Wang}).

Signal/image processing is the discipline that develops and studies the techniques for processing, analyzing and interpreting signals. Among the types of operations possible on these signals / images, we can denote control, filtering, compression and transmission of data, noise reduction, deconvolution, prediction, identification, classification, etc.

Although this discipline has its origin in the engineering sciences (particularly electronics and automation), today it makes extensive use of many areas of mathematics, such as signal theory, stochastic processes, vector spaces and linear algebra and applied mathematics, in particular information theory, optimization or numerical analysis. In this context, linear transformations have always played a very important role, and among these, the best known and oldest is of course the Fourier transform. See for example \cite{Bacchelli-Cotronei-Sauer,Mallat,Kotas,Wang}\,.  

Images and signals especially biomedical ones as we said previously continue to be an important task for scientific researches from both theoretical and practical point of view. Some new pandemics appeared recently have for example to be understood as they constitute a real and strong threatening of the humanity such as SRAS, H2N2 and the new Coronavirus. One of the powerful tools in such topics is wavelet theory which has been proved to be challenging since its discovery. Indeed, the notion of wavelets has known since its discovery a great growing up especially in the field of signal and image processing. Recently, a step forward has been conducted to extend wavelets to multiwavelets to cover some multidimensional cases of signals. In the present work we propose to improve multiwavelet notion by adopting more general families of multiwavelets explicitly constructed using independent components for multidimensional scaling functions leading to a new more flexible 2-scale relation. Readers may also consult \cite{BenMabroukBenAbdalllahDhifaoui,Arfaoui,Jallouli1,Jallouli2,Jallouli3,Jallouli4,Zemni1} for more applications of wavelets and multiwavelets on biosignals.

Multimedia documents constitute also a category of applications in signal and image processing. They also present essential tools in the fields of biomedical, satellite and astronomical imagery, film production, and industrial computing as solutions to problems related to the transmission of confidential data, big, fuzzy, missing and cloud data, cryptography, watermarking, steganography, ... In this context, the image undergoes transformations and the resulting image is then tested by a detection process which extracts the mark or points out its presence. In watermarking for example, some methods operate in the spatial or transformed domain and others use hybrid techniques. Methods based on transformations consist of transforming the signal of the space-time domain to another domain (for example Fourier, wavelets, ...). It is done by inserting the mark in some coefficients in this field. Then the inverse transformation is performed to return to the space-time domain. Concerning the frequency transform, the insertion of the mark in the low frequencies generally provides good robustness but induces distortions in the time domain. On the other hand, the insertion in the high frequency components does not keep the quality and moreover it makes the mark fragile to the attacks. This was a motivation for researchers to develop and ensure a compromise between the robustness and invisibility of the transform. See \cite{Arfaouietal,Bacchelli-Cotronei-Sauer,Mallat,Kotas,Wang}\,.

Over the past few years, there has been a renewed interest in multiresolution representations, surface filtering, contour detection and information retrieval. When it is sought to analyze an image, it is very common to establish, explicitly or implicitly, a time-frequency representation of it. The Fourier transform is not the appropriate tool to carry out this analysis since it masks the temporal evolution of the signal. In the eighteenth, wavelet theory has been rediscovered and proved to be a powerful tool in signal/image processing. Since its discovery, wavelets were related to the theory of signals/images as applied during the study of a reflected oil extracting signal. The reflected signal is not stationary! Next, wavelet transform has been the subject of numerous studies in signal processing and geometrical computing. Indeed, most of the signals of the real world are not stationary, and it is just in the evolution of their characteristics (statistics, frequency, temporal, spatial) that resides the essential information that they contain. Local signals and images are exemplary. In this context, wavelet transforms provide information about the frequency content while preserving the localization in time in order to obtain a time-frequency representation or a space-scale of the signal. Unlike the Fourier transform, the wavelet transform and its extensions provide interesting solutions in this context. Approximations of signals are obtained as results of a convolution with a scaling function (a low-pass filter) and a wavelet function, and then reducing the number of points used in the process. The principle idea is to iterate this process and transform the current approximation into a new approximation with fewer points for the representation. We obtain a temporal as well as a frequency decomposition of the source object. It is well known that the frequency decomposition of a signal is interesting for the analysis of the different levels of detail present in the signal. It also applies to filtering, compression and progressive transmission. See \cite{Arfaouietal,Bacchelli-Cotronei-Sauer,Daubechies,Mallat,Kotas,Wang}.  

In the present work we propose to serve from explicit multiwavelets already introduced in \cite{Zemni1} and next applied in \cite{Zemni2} to improve firstly the theoretical findings and in modeling biomedical signals. The basic idea is consists in a simple change in the well-known 2-scale relation by writing it in a vector form. This makes almost all existing constructions of multiwavelets to look-like as modified representations of the same original wavelets. See \cite{AlMahamdya-Riley,Alramahietal,Alwan,Attakitmongcol-Hardin-Wilkes,Massopust-Ruch-VanFleet,Riederetal,Turcajova}. In our work, based on the well-known wavelets of Haar and Faber-Schauder we developed a simple variant of multiwavelets that are not issued from one source as existing ones. Haar and Schauder explicit functions are applied in our case. This choice permits exact computations of necessary coefficients applied in the processing. They also permit to reduce the number of such coefficients and obtain the next generations recursively. However, we recall that other examples of multiwavelets may be also obtained even explicitly by applying other scaling functions and/or wavelet mothers different from the present case. Some interesting cases may be found in \cite{Arfaoui1,Arfaou2}. See  also  \cite{Bui-Chen,Huangetal,Iyer,Kessler,Liangetal,Ruedin1,Selesnick1,Selesnick2,Selesnick3,Tham-Shen-Lee-Tan,Vehel-Aldroubi,Xia,Xia-Jiang,Xia-Geronimo-Hardin-Suter} for more methods and applications.

Next, to show the performance of our extension, some experimentations will be developed. A first one deals with the development of a Fourier type mode to show how fast are algorithms based on the new variant of multiwavelets. A second experimentation will be concerned with the well known ECG signals. Recall that the technique of ECG consists of measuring the differences of potentials due to the dipole field, at different points of the body. It produces a graphical representation of the heart electrical activity. The main problem for such a signal is the presence of noise. Hence, a denoising step has to be conducted using our new multiwavelets to lead next to a good analysis. See \cite{Kotas,Wang,AlMahamdya-Riley} for some existing methods. The last experimentation is concerned with the processing of a Coronavirus strain for an associated membrane protein signal. We propose to develop a wavelet analysis of an isolated or purified strain of human coronavirus associated with SARS already recorded and studied in \cite{VanDerWerf}. Precisely, we intend to conduct a reconstruction process and to localize membrane helices of the strain based on the hydrophobic character of the amino acids constituting the proteins' series associated to such a strain and issued from the well known Kyte-Dolittle method \cite{KyteJ}.

The present paper is organized as follows. The next section is devoted to the review of wavelet theory. Section 3 is devoted to the development of multi-wavelet concepts in order to provide a Haar-Schauder multi-wavelet and its associated filters. Instead of introducing the multiwavelet scaling function as the vector composed of the translated copies of the same single scaling function appearing in the 2-scale relation a new concept of multiwavelet scaling function is introduced based on finitely many possibly independent scaling functions components. It looks like a system of many cameras working simultaneously and independently in each direction. In section 4, some experiments have been developed to show the performance of multiwavelets against wavelets for both the rapidity of algorithms and biosignals processing. An ECG signal and a proteins' strain issued from a coronavirus case are considered.
\section{Brief review on wavelets}
A wavelet may be defined simply and especially for a non mathematicians community as a short wave function and which has major difference from Fourier sine and cosine by its ability of being localized in time-frequency and/or time-space. Wavelet analysis of signals is based also on the so-called wavelet transform which is a convolution of the analyzed signal with copies of a source function called mother wavelet. Wavelets,differently from Fourier modes, are not necessarily periodic, they may be also compactly supported. 

In mathematics, a mother wavelet $\psi$ is a square-integrable function with enough vanishing moments (oscillating) with necessary zero mean. The copies applied next in the signal analysis are issued from the mother wavelet by acting the so-called real affine group $G_+$, also called the $ax + b$ group, consisting of transformations of $\mathbb{R}$ of the type $x\longmapsto gx=ax+b$, $x\in\mathbb{R}$, where $a > 0$, $b\in\mathbb{R}$ equipped by the multiplication rule 
$$
g_lg_2=(b_1+a_1b_2,a_1a_2),\;g_1=(a_1,b_1),\,g_2=(a_2,b_2)\in G_+.
$$
Denote $T_bf(x) = f (x - b)$, $D_af(x) =\sqrt{a} f (ax)$, $a> 0$ and $(U(a,b)f)=D_{a^{-1}}T_bf$. It holds that 
$$
(U(a,b)f)(x)=\dfrac{1}{\sqrt{a}}f(\dfrac{x-b}{a}).
$$
The wavelet transform of $f$ at the scale $a>0$ and the position $b\in\mathbb{R}$ is
\begin{equation}\label{CWT-1}
	C_f(a,b)=\langle f,U(a,b)\psi\rangle=\langle f,D_{a^{-1}}T_b\psi\rangle.
\end{equation}
This transform is invertible and its inverse is 
\begin{equation}\label{InverseCWT-1}
	f(\bullet)=\dfrac{1}{\mathcal{A}_{\psi}}\langle C_f(a,b),U(a,b)\psi(\bullet)\rangle_{d_\mu(a,b)}
\end{equation}
where $\mathcal{A}_{\psi}$ is the admissibility constant due to the mother wavelet ${\psi}$ expressed as
\begin{equation}\label{admissibility}
	\mathcal{A}_{\psi}=\displaystyle\int_{\mathbb{R}}\dfrac{|\widehat{\psi}(\xi)|^2}{|\xi|}{d\xi}<\infty.
\end{equation}
(See \cite{Arfaouietal,Daubechies,Mallat}).

In the sequel we will denote 
\begin{equation}
	{\psi}_{a,b}(x)=(U(a,b)\psi)(x)=\dfrac{1}{\sqrt{a}}\psi{}(\dfrac{x-b}{a}).
\end{equation}
The wavelet processing of signals is based on their wavelet transform which may be continuous or discrete. Given a finite energy signal $F$, $a>0$ known as the scale and $b\in\mathbb{R}$ known as the position, the CWT of $F$ is (at the scale $a$ and the position $b$) is as introduced in (\ref{CWT-1}) and recalled here-after
\begin{equation}
	C_F(a,b)=\displaystyle\int_{-\infty}^{+\infty}F(t) {\psi}_{a,b}(t)dt.
\end{equation}
The analyzed signal $F$ may be reconstructed using the inverse transform as in (\ref{InverseCWT-1}) as
\begin{equation}\label{reconstructionformule}
	F(t)=\displaystyle\dfrac{1}{\mathcal{A}_\psi}\displaystyle\int_{-\infty}^{+\infty}C_F(a,b){\psi}_{a,b}(t)\displaystyle\dfrac{dadb}{a^2}.
\end{equation}
See \cite{Arfaouietal,Daubechies}. 

A restrictive version of the CWT is the so-called discrete wavelet transform called also wavelet coefficient which is the restriction of the continuous form to a discrete set of the scale and the position parameters. In fact there is no essential difference between the discrete grids used and the most commonly used one is the dyadic grid constituted by $a=2^{-j}$ and $b=k2^{-j}$, $j,k\in\mathbb{Z}$. The copy $\psi_{a,b}$ becomes is this case
\begin{equation}
	\psi_{j,k}(t)=2^{j/2}\psi(2^jt-k)
\end{equation}
and the discrete wavelet transform (DWT) (or the wavelet coefficient) will be
\begin{equation}
	d_{j,k}(F)=\displaystyle\int_{-\infty}^{+\infty}F(t)\psi_{j,k}(t)dt.
\end{equation}
These coefficients are also known in wavelet theory as the detail coefficients at the level $j$ and the position $k$. It holds also in wavelet theory that $(\psi_{j,k})_{j,k\in\mathbb{Z}}$ constitutes an orthonormal basis of $L^2(\mathbb{R})$ and consequently any element $F$ may be decomposed in a series 
\begin{equation}\label{SerieDondelettesdeS1}
	F=\sum_{j,k}d_{j,k}(F)\psi_{j,k}
\end{equation}
known as the wavelet series of $F$ and which replaces the reconstruction formula (\ref{reconstructionformule}) in the discrete form. 

This decomposition into an orthogonal-wise components series leads to a functional framework associated to the ,other wavelet $\psi$ known as the multiresolution analysis (MRA). Indeed, let for $j\in\mathbb{Z}$, $W_j=spann(\psi_{j,k}\;;\;k\in\mathbb{Z})$ known as the detail spaces and $V_j=\displaystyle\oplus_{l\leq j}W_l$ called approximation spaces. There exists a source function $\varphi$ known as the scaling function or the father wavelet satisfying  $V_j=spann(\varphi_{j,k}\;;\;k\in\mathbb{Z})$, where the $\varphi_{j,k}$'s are defined similarly to the $\psi_{j,k}$. The father and mother wavelets are related by the so-called 2-scale relation stating that
\begin{equation}\label{equation2echelle}
	\varphi=\displaystyle\sum_{k\in\mathbb{Z}}h_k\varphi_{1,k}\;\;\mbox{and}\;\;
	\psi=\displaystyle\sum_{k\in\mathbb{Z}}g_k\varphi_{1,k}
\end{equation}
where
\begin{equation}\label{psiapartirdephi}
	h_k=\displaystyle\int_{-\infty}^{+\infty}\varphi(t)\varphi_{1,k}(t)dt\;\;\mbox{and}\;\;g_k=(-1)^kh_{1-k}.
\end{equation}
See \cite{Daubechies,Mallat} for more details. These relations permit to compute the wavelet coefficients from level to level. Indeed, denote
$$
a_{j,k}(F)=\displaystyle\int_{-\infty}^{+\infty}F(t)\varphi_{j,k}(t)dt
$$
known as the approximation or the scaling coefficient of $F$ at the level $j$ and the position $k$, we have
\begin{equation}
	a_{j,k}(F)=\displaystyle\sum_{l\in\mathbb{Z}}h_{l}a_{j+1,l+2k}(F)
\end{equation}
and
\begin{equation}
	d_{j,k}(F)=\displaystyle\sum_{l\in\mathbb{Z}}g_{l}a_{j+1,l+2k}(F).
\end{equation}
This means that the decomposition at the level $j$ may be deduced from the level $(j+1)$ by means of the filters $H=(h_k)_k$ (discrete wavelet low-pass filter) and $G=(g_k)_k$ (discrete wavelet high-pass filter). Similarly, we have an inverse scheme stating that 
\begin{equation}\label{AlgReconstruction}
	a_{j+1,k}(F)=\displaystyle\sum_{l}h_{l-2k}a_{j,l}(F)
	+\displaystyle\sum_{l}g_{l-2k}D_{j,l}(F).
\end{equation}
For backgrounds on wavelet filters, the readers may refer to \cite{Arfaouietal,Daubechies,Mallat}.
\section{Generalized multiwavelet analysis}
Multwavelets have been introduced since the early 1990s as another view of wavelets permitting to re-write wavelet analysis in a vector form to reduce may be mathematical formulations. It resembles in some sense to the reduction of higher order differential equqtions into first order ones by considering the vector $X=(y,y',y'',\dots,y^{(n)})$ where $n\in\mathbb{N}$ is an integer constituting the order of the original differential equation in $y$ and where $y',y'',\dots,y^{(n)}$ are the derivatives of $y$ to such an order. The major existing multiwavelet constructions consider the vector $\Phi=(\varphi(.),\varphi(.-1),\dots,\varphi(.-N))$ where $N$ is the length of the filters $H$ and $G$. 

This view of wavelets has even though some advantages such as short supports, smoothness, accuracy, symmetry and orthogonality. Moreover, as noticed in \cite{Zemni1,Zemni2}, discrete multiwavelets may require pre-processing and post-processing steps. These facts themselves constituted main motivations behind the study developed in \cite{Zemni1,Zemni2} and continued in the present paper. See also 
\cite{Efromovich,Fowleretal,Johnson,Lebrun-Vetterli1,Lebrun-Vetterli2,Ruedin2,Shen-Tan,Stacey-Blyth,Strela-Heller-Strang-Topiwala-Heil,Yoganand-Mohan}. 

In the present paper, we propose to continue in exploiting more the construction of multiwavelets as noticed in \cite{Zemni1,Zemni2} by considering a vector-valued scaling function 
$\Phi=\left(\varphi_1,\varphi_2,\dots,\varphi_N\right)^T$ ($^T$ is the transpose), $N\in\mathbb{N}$ fixed where the components $\varphi_i$, $i=1,2,...,N$ are not translations of the same function as in the most existing case. This leads to a matrix-vector 2-scale relation 
\begin{equation}\label{2ScaleRelationMultiwavelets}
	\Phi=\displaystyle\sum_{k}H_k\Phi_{1,k}
\end{equation}
where in this way the $H_k$'s are $(N,N)$-matrices, $H_k=\bigl(h_{i,j}\bigr)_{1\leq i,j\leq N}$. Similarly, the mother multiwavelet will satisfy a scale relation of the form
\begin{equation}\label{MultiwaveletPsiFromPhi}
	\Psi=\displaystyle\sum_{k}G_k\Phi_{1,k}
\end{equation}
where the $G_k$'s are also $(N,N)$-matrices, $G_k=\bigl(g_{i,j}\bigr)_{1\leq i,j\leq N}$.
\begin{definition}
	The sequences of matrices $H=(H_k)_k$ and $G=(G_k)_k$ are called the discrete high pass and discrete low pass multi-filters respectively.
\end{definition}
In the literature review on multiwavelets, there are few developments. So, complete and full exposition of multiwavelets theory still needs to be developed. The only reference in this direction is \cite{Keinert}. This is one motivation among previous ones letting us to develop the present work. The choice of mother multiwavelets is also strongly related to the ability and flexibility in conducting experiments Readers may refer to \cite{Attakitmongcol-Hardin-Wilkes,Brazile,Hardin-Roach,Keinert,Stankovicetal,Zhang-Davidson-Luo-Wong,Wang}

In the sequel, we fix the multiwavelet order $N=2$. Let $\varphi_1(x)=\chi_{[0,1[}(x)$ be the Haar scaling function and $\varphi_2(x)=(1-|x|)\chi_{[-1,1[}(x)$ be the Schauder scaling function. Denote next
$\Phi=\left(\varphi_1,\varphi_2\right)^T$. Simple calculus yield that $H_k=0$ whenever $|k|\geq2$ and  
\begin{equation}
	\Phi=H_{-1}\Phi_{1,-1}+H_{0}\Phi_{1,0}+H_{1}\Phi_{1,1}
\end{equation}
where
\begin{equation}\label{eq11}
	H_{-1}=H_{1}=\displaystyle\frac{1}{\sqrt2}
	\left(\begin{array}{lll}
		0&0\\ 0&1/2\end{array}\right),
	\;H_{0}=\displaystyle\frac{1}{\sqrt2}
	\left(\begin{array}{lll}
		1&0\\ 0&1\end{array}\right).
\end{equation}
Thus, the mother multiwavelet is
\begin{equation}\label{PsiMultiwaveletandPhiRelation}
	\Psi=\displaystyle\sum_lG_l\Phi_{1,l},\;\;G_l=(-1)^lH_{1-l}.
\end{equation}
The Haar-Schauder multiwavelet processing (decomposition/reconstruction) of a signal $F$ consists as in all wavelet processing in estimating the corresponding coefficients of the signal by means of the multiwavelet copies. So consider for $r\in\mathbb{N}$ fixed known as the order or the dimension of the signal and write $F=\left(F_1,F_2,\dots,F_r\right)^T$. Denote also s in single wavelet analysis 
$A_{j,k}(F)$ and $D_{j,k}(F)$ the approximation and the detail coefficients of $F$ relatively to the Haar-Schauder multiwavelets at the level $j$ and the position $k$. The signal $F$ may be decomposed as a sum
$$
F=A_0+D_0
$$
where $F_0$ is  
\begin{equation}\label{S00}
	A_0=\displaystyle\sum_lA_{0,l}(F)\Phi_{0,l}
\end{equation}
and 
\begin{equation}\label{DS00}
	D_0=\displaystyle\sum_lD_{0,l}(F)\Psi_{0,l}.
\end{equation}
The components $A_0$ and $D_0$ are known as the approximation and the detail components of $F$ at the level $0$. The coefficients $A_{0,l}(F)$ and $D_{0,l}(F)$ are $(r,2)$-matrices. As in the case of single wavelet theory, we obtain here a MRA associated to the multiwavelet by considering as approximation space $V_0$ the closure of vector space spanned by the $\Phi_{0,l}$ and as detail space at the level 0 the one spanned by $\Psi_{0,l}$, $l\in\mathbb{Z}$. As a consequence, we obtain multiwavelet algorithms stating that 
\begin{equation}\label{ReconstAlg1}
	A_{1,s}(F)=\sum_l\left[A_{0,l}(F)H_{s-2l}+D_{0,l}(F)G_{s-2l}\right],
\end{equation}
\begin{equation}\label{DecompAlg1-1}
	A_{0,s}(F)=\sum_l\,A_{1,l}(F)H_{l-2s}
\end{equation}
and 
\begin{equation}\label{DecompAlg1-2}
	D_{0,s}(F)=\sum_l\,A_{1,l}(F)G_{l-2s}.
\end{equation}
To resume, the new general concept will cover some disadvantages of many existing multiwavelets theory where the scaling multiwavelet function is constructed by taking the well known 2-scale relation in single wavelet theory and introducing the multiwavelet scaling function as the vector composed of the translated copies of the single source scaling function appearing in the 2-scale relation. More precisely, let $\varphi$ be a scaling function satisfying an associated 2-scale relation
$$
\varphi(x)=\displaystyle\sum_{k=0}^{L-1}h_k\varphi(2x-k),
$$
the associated multiwavelet is $\Phi(.)=(\varphi(.),\varphi(.-1),\dots,\varphi(.-L+1))$ where $L$ is the filter lenth. This is good in some way as it lokks like a system of $L$ surveillance systems in each direction, but which are indentical or having the same mechanism. However, it will be best and more efficient to install different mechanisms' cameras and thus get a whole system of surveillance $\Phi(.)=(\varphi_1,\varphi_2,\dots,\varphi_K)$ with a number of directional-wise cameras with different filters,independent and working simultaneouslyto compose awhole image. In the present work one of our aims is to apply the last mechanism of multiple different cameras.
\section{Experimentation}
\subsection{Development of a Fourier mode}
In this section we aim to develop the wavelet analysis of a special example of signals consisting of the well known $2\pi$-periodic Fourier mode $F(t)=\sin(t)$, $t\in[0,2\pi]$. Recall that the decomposition de $F$ at the level $J\in\mathbb{N}$ is expressed as
\begin{equation}
	F=\displaystyle\sum_kA_{J,k}(F)\varphi_{J,k}+\displaystyle\sum_{j\geq J}\displaystyle\sum_kD_{j,k}(F)\psi_{j,k}.
\end{equation}
For a choice of $J=1$, the approximation part becomes
\begin{equation}
	A_1=\displaystyle\sum_kA_{1,k}(F)\varphi_{1,k}.
\end{equation}
Recall now that
\begin{equation}
	A_{1,k}(F)=\displaystyle\int_{(k-1)/2}^{(k+1)/2}\sin(t)\varphi_{1,k}(t)\chi_{[0,2\pi[}(t)dt.
\end{equation}
We now compute the values of the position parameter $k$ for which the intersection of supports $[\frac{k-1}{2},\frac{k+1}{2}[\cap[0,2\pi[\not=\emptyset$ which yields that $0\leq k\leq[4\pi]$. 

We next compute the Normalized Average Quadratic Error (NAQE) to show the performance of the approximation computed on a grid of $N$ points $t_i$ in $[0,2\pi]$,
\begin{equation}\label{NAQE-Error}
	NAQE_{J,N}(A_1,F)=\displaystyle\frac{\displaystyle\sum_{i=1}^{N}(A_J(t_i)-F(t_i))^2}
	{\displaystyle\sum_{i=1}^{N}(F(t_i))^2}.
\end{equation}
For a number $N=50$ and $J=1$, we get an error 
$$
NAQE= 0.0012.
$$
The following figure (Fig.7) illustrates the signal $F$ and its approximation $A_1$.
\begin{figure}[!hbp]
	\begin{center}
		\includegraphics[scale=0.60]{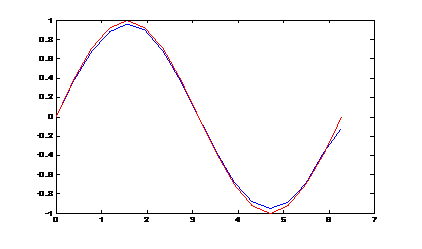}
		\caption{$F$ (red) and its approximation $A_1$ (blue).}
		\label{fetA}
	\end{center}
\end{figure}

Next, to show the role of the projections of the signal $F$ on the detail spaces we compute the DWT of $F$ already with $J=1$. This will illustrate the dynamics of $F$. Similarly to the approximation case, it remains for the position parameter $k$ the values $-1$, $0$, $1$, $\dots$, $13$. Recall that the support of $\psi_{1,k}$ is $[\frac{k-1/2}{2},\frac{k+1/2}{2}]$. Thus, to get the detail component $D_1$ of the signal $F$ in the detail space $W_1$ we have to compute $D_{1,k}(F)$ for $k\in\{-1,0,1,\dots,13\}$. 

Next, denote $F_1=A_1+D_1$. To illustrate the closeness of $F_1$ to the original signal $F$, we compute as previously the NAQE on a grid of $N$ points $t_i$ in $[0,2\pi]$,
\begin{equation}\label{NAQE-Error-1}
	NAQE_N(F_1,F)=\displaystyle\frac{\displaystyle\sum_{i=1}^{N}(F_1(t_i)-F(t_i))^2}{\displaystyle\sum_{i=1}^{N}(F(t_i))^2}.
\end{equation}
For a number $N=50$, we get an error  
$$
NAQE= 0.00118.
$$
The following figure (Fig. 8) illustrates the signal $F$ and its approximation $F_1$.
\begin{figure}[!hbp]
	\begin{center}
		\includegraphics[scale=0.60]{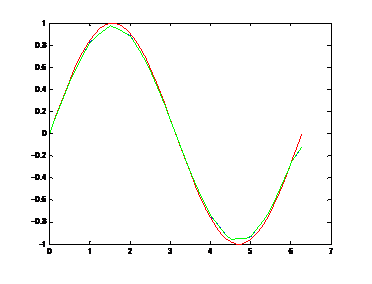}
		\caption{$F$ (red) and $F_1$ (green).}
		\label{psischauder}
	\end{center}
\end{figure}

Similarly, we may compute for $J\in\mathbb{N}$ the approximation
\begin{equation}\label{JlevelDecomp}
	F_J=A_J+D_1+D_2+\dots+D_J.
\end{equation}
To illustrate the closeness of these approximations to the original signal $F$, we compute the Normalized Average Quadratic Error (NAQE) on a grid of $N$ points $t_i$ in $[0,2\pi]$. For a number $N=50$ and $J=1$, we get the following error estimates (Table \ref{comparisontable}).
\begin{table}[h!]
	\begin{center}
		\begin{tabular}{||c|c|c|c||}
			\hline\hline
			Method 1 (Ref. \cite{Brazile})&Bi-filters $J=1$&Bi-filters $J=2$&Bi-filters $J=3$\\
			\hline
			12.$10^{-3}$&11,8.$10^{-3}$&2.$10^{-5}$&2,4.$10^{-6}$\\
			\hline\hline
		\end{tabular}
		\medskip
		\caption{Error Estimates.}\label{comparisontable}
	\end{center}
\end{table}

Table \ref{comparisontable} summarizes the results of comparisons with the existing method developed in \cite{Brazile} and the bi-filters based method developed here. We found that NAQE obtained by bi-filters is smaller than the existing one. On the other hand, it is remarkable that the greater $J$ increases the error decreases.

Next, in order to show more the performance of the new method we proposed to evaluate the running time of algorithms due to each method. We thus provided a comparison relatively to the time execution algorithms for the methods applied for the same Fourier mode signal. For $N=10$ and $J =1$, we obtained the following table (Table \ref{comparisontable1}).
\begin{table}[h!]
	\begin{center}
		\begin{tabular}{||c|c|c|c||}
			\hline\hline
			The method&NAQE&Running Time\\
			\hline
			Schauder Wavelet&0,0086&123,2\,s\\
			\hline
			Schauder Filters&0,0092&73,03\aa,s\\
			\hline
			Haar-Schauder Muliwavelet&0,0033&32,97\,s\\
			\hline
			Haar-Schauder Multiwavelet Filters&0,0003&16,6\,s\\
			\hline\hline
		\end{tabular}
		\medskip\caption{Time execution.}\label{comparisontable1}
	\end{center}
\end{table}

Table \ref{comparisontable1} shows a comparison for both the NAQE error and the execution time between the approximation obtained by the use of the Schauder wavelet, Schauder Filters, Haar-Schauder Multiwavelet and Haar-Schauder Multiwavelet Filters. First, by comparing the NAQE and the execution time for the methods based on the single Schauder wavelet and Schauder filters we noticed that the NAQE relative to both of them are not enough different. Besides, the second one yields a faster convergent algorithm. Next, applying Haar-Schauder multiwavelets results in more efficient approach. Similarly to the single case, the new Haar-Schauder multiwavelet filters result in a best error and a best running time. This shows the performance of the new multiwavelet approach. Finally, our work proves among the efficiency of multiwavelet approaches, that using different wavelet cells in the multiwavelet black boxes is more performant than applying the classical approach. Recall that this latter is on re-writing the 2-scale relation and thus re-writing the whole signal in a different way by decomposing it in different bi-signals, which may affect the originality of the signal processed.
\subsection{ECG signal processing}
ECG signals are graphical representations of the heart electrical activity due to the variations of electric potential of the specialized cells in the contraction (myocytes) and specialized cells in the automatism and the conduction of the influxes. ECG can highlight various cardiac abnormalities and has an important place in cardiology diagnostic tests, as for coronary heart disease. We refer to the MIT-BIH Arrhythmia data basis for the application develiped in this part. 

The present ECG signal processing by means of the Haar-Schauder multiwavelet yields for each level of decomposition $J\geq1$ a discrete positions grid $0\leq k\leq 10.2^J$. 

\begin{figure}[h]
	\begin{center}
		\begin{tikzpicture}
			\begin{scope}[xscale=1.5,yscale=1.5]
				\node (N0) at (0,7) [rectangle,draw]{The original signal};
				\node (N11) at (-3,6) [rectangle,draw]{Project on $V_J$};
				\node (N12) at (-1,6) [rectangle,draw]{Project on $W_J$};
				\node (N13) at (1,6) [rectangle,draw]{Project on $\widetilde{V}_J$};
				\node (N14) at (3,6) [rectangle,draw]{Project on $\widetilde{W}_J$};
				\node (N21) at (-2,5) [rectangle,draw]{${a}_J$};
				\node (N22) at (-1,5) [rectangle,draw]{${d}_J$};
				\node (N23) at (1,5) [rectangle,draw]{$\widetilde{a}_J$};
				\node (N24) at (2,5) [rectangle,draw]{$\widetilde{d}_J$};
				\node (NF) at (0,4) [rectangle,draw]{The reconstructed signal};
				\draw[->] (N0) -- (N11);
				\draw[->] (N0) -- (N12);
				\draw[->] (N0) -- (N13);
				\draw[->] (N0) -- (N14);
				\draw[->] (N11) -- (N21);
				\draw[->] (N12) -- (N22);
				\draw[->] (N13) -- (N23);
				\draw[->] (N14) -- (N24);
				\draw[->] (N21) -- (NF);
				\draw[->] (N22) -- (NF);
				\draw[->] (N23) -- (NF);
				\draw[->] (N24) -- (NF);
			\end{scope}
		\end{tikzpicture}
		\caption{The Haar-Schauder multiwavelet principle.}
		\label{AnalyseFigure2}
	\end{center}
\end{figure}
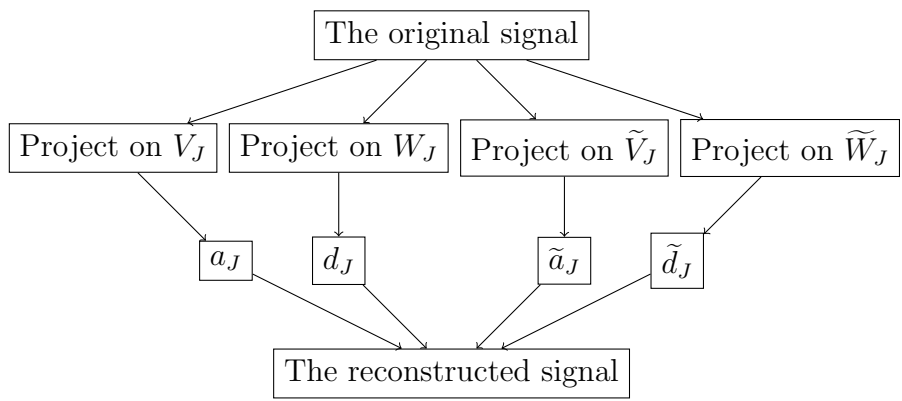

Next, our idea consists in using the Haar-Schauder multiwavelet for the ECG signal processing as a type of simultaneous loops to guarantee the maximum information carried in such a signal. The first loop consists in applying a filtering of the signal by means on one of the components of the Haar-Schauder multiwavelet (Haar for example) and next apply the second one to denoise more the obtained filtered sub-signal. This raises an interesting question about the use of independent components in the definition of the multiwavelet analysis source functions $\Phi$ and $\Psi$. This filtering concept could not be realised by using multiwavelets with non-separable variables and/or dependent components. So, the idea is a double (multiple in general) surveillance cameras system that is used to detect best strange objects. The following diagram (Figure \ref{AnalyseFigure2}) shows the decomposition steps of an ECG signal using Haar-Schauder multiwavelet.

We now explain mathematically the principle of Haar-Schauder multiwavelet processing. So, denote $\varphi$ and $\widetilde{\varphi}$ the Haar and Faber-Schauder scaling functions respectively and the associated mother wavelets $\psi$ and $\widetilde{\psi}$. For a level $J$ denote $a_{J}$ and $\widetilde{a_{J}}$ the approximations at the level $J$ due to Haar and Faber-Schauder MRA respectively and similarly $d_{J}$ and $\widetilde{d_{J}}$ the projections on the detail spaces due to Haar and Faber-Schauder MRA respectively. We get the multiwavelet decomposition of the ECG signal at the level $J$ as 
$$
A_J=a_{J}+\widetilde{a_{J}}+\sum_{j}^J{d_{j}}+\sum_{j}^J{\widetilde{d_{j}}}.
$$
Using the independence between the components of the multiwavelet, the principle applied here means that the final decomposition is a superposition of two decompositions on two approximation spaces and two detail spaces for each level included in the modeling. In this case, the risk of losing the information decreases compared with classical wavelet processing. The reconstruction by multiwavelet will be more efficient. Moreover, it is worth to recall here that there is no essential difference between being simultaneous or consecutive  the application of the two components of the multiwavelet. Such a problem may be of great importance when the components are dependent or depending on non-separable variables. 

Table \ref{comparisontable3} resumes the accuracy of the present method against previous ones by means of the so-called Normalized Average Quadratic Error (NAQE) as in (\ref{NAQE-Error}) or (\ref{NAQE-Error-1}).
\begin{table}[h]
	\begin{center}
		\begin{tabular}{|c|c|c|c|c|}
			\hline
			NAQE&$J=1$&$J=2$&$J=3$&$J=4$\\
			\hline
			Haar wavelet&0.0012&8.7 $10^{-4}$&5.46 $10^{-4}$&5.01 $10^{-4}$\\
			\hline
			Schauder wavelet&0.0014&8.95 $10^{-4}$&6.18 $10^{-4}$&3.7 $10^{-4}$ \\
			\hline
			Haar-Schauder multiwavelet&9.8 $10^{-4}$&7.4 $10^{-4}$&4.37 $10^{-4}$&1.09 $10^{-4}$\\
			\hline
		\end{tabular}
		\caption{Relative NAQE estimates for ECG signal.}
		\label{comparisontable3}
	\end{center}
\end{table}

\begin{figure}[h]
	\begin{center}
		\includegraphics[scale=0.50]{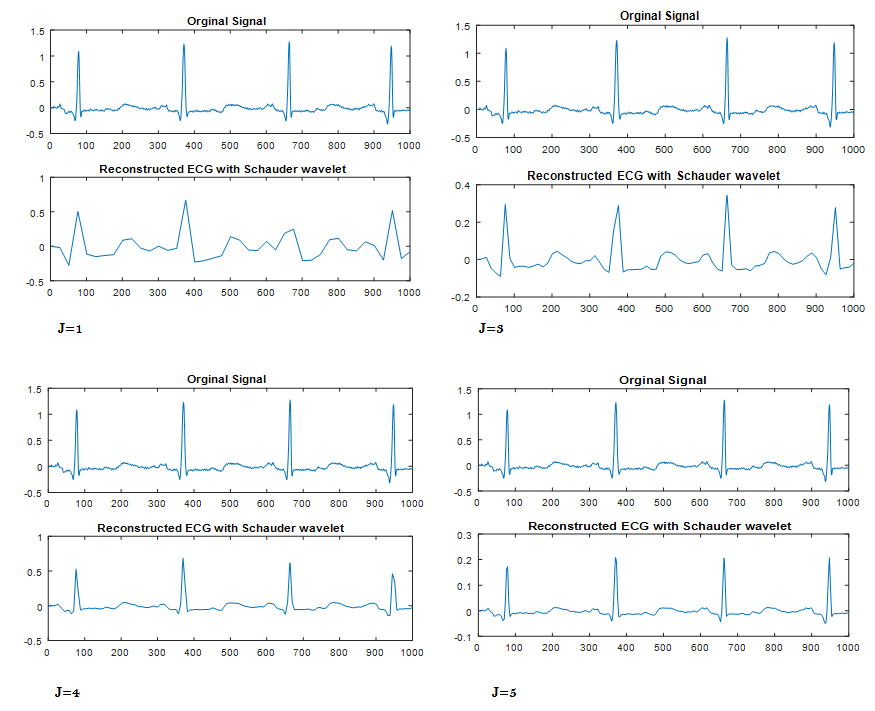}
		\caption{Reconsruction of the ECG signal by Schauder wavelet}
		\label{AnalyseFigurewsh}
	\end{center}
\end{figure}

\begin{figure}[h]
	\begin{center}
		\includegraphics[scale=0.50]{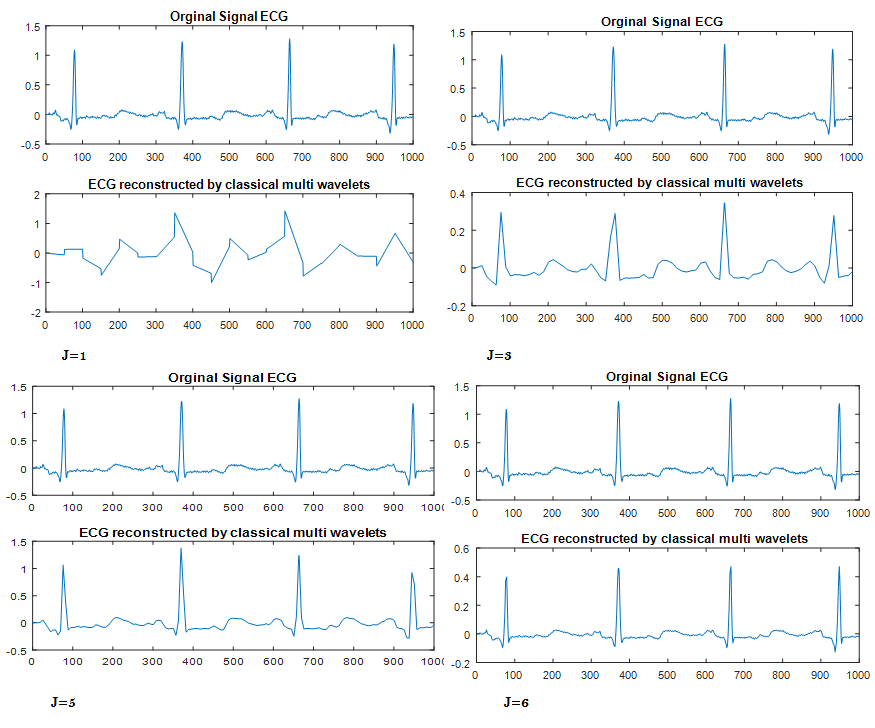}
		\caption{Reconsruction of the ECG signal by Haar wavelet}
		\label{AnalyseFigurewha}
	\end{center}
\end{figure}

\begin{figure}[h]
	\begin{center}
		\includegraphics[scale=0.450]{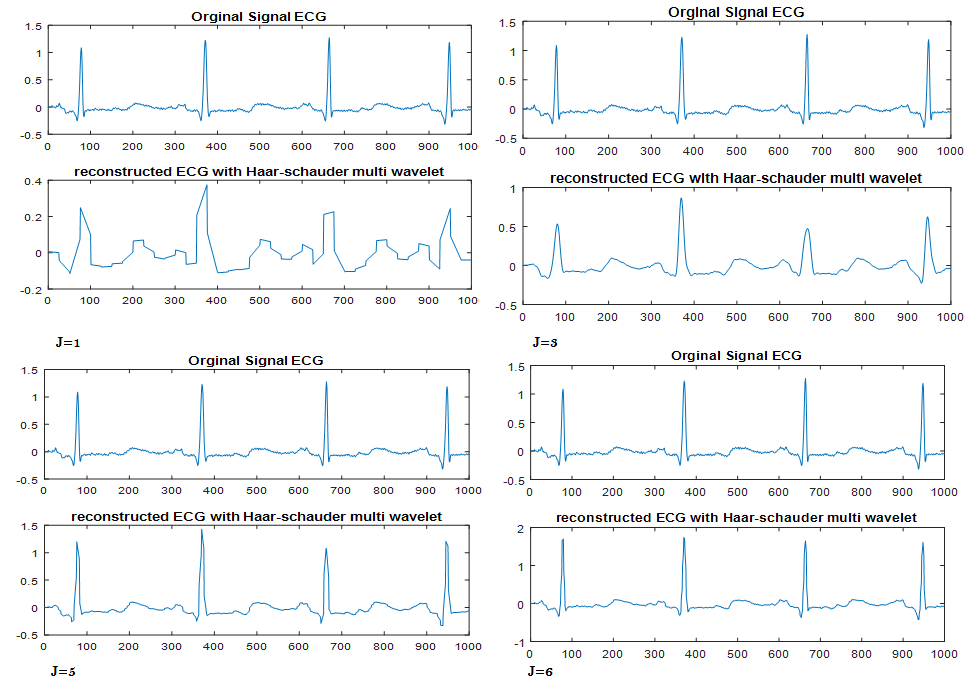}
		\caption{Reconsruction of the ECG signal by the Haar-Schauder multiwavelet}
		\label{AnalyseFigurewmw}
	\end{center}
\end{figure}

\begin{figure}[h]
	\begin{center}
		\includegraphics[scale=0.60]{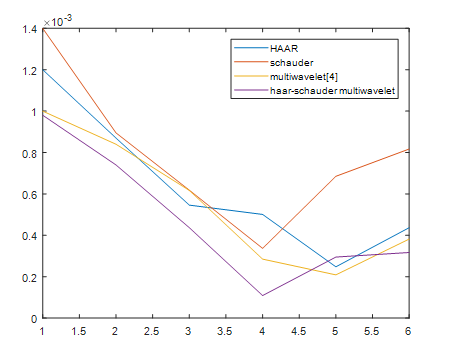}
		\caption{Error estimates relatively to the decomposition level $J$ for ECG signal.}
		\label{AnalyseFigurewmw2}
	\end{center}
\end{figure}

We notice easily from Table \ref{comparisontable3} that the Haar-Schauder multiwavelet processing results in more accurate error of closeness NAQE obtained for the best estimates at a level of decomposition $J=4$. This proves also that the multiwavelet processing did not necessitate a higher order of decomposition to reach a good error. Besides, Figures \ref{AnalyseFigurewsh}, \ref{AnalyseFigurewha} and \ref{AnalyseFigurewmw}
illustrate the processing of the ECG signal using Haar wavelet, Schauder wavelet and Haar-Schauder multiwavelets and confirm more the efficiency and the performance of the multiwavelet principle.\\
\\
\\
\\
To finish with the ECG multiiwavelet processing we plotted in Figure \ref{AnalyseFigurewmw2} the evolution of the normalized quadratic error with the level of resolution $J$, for Haar wavelet, Schauder wavelet and Haar-Schauder multiwavelets. The graph shows easily the efficiency of wavelet processing in general and more efficiently the dominance of the new multiwavelet against the single wavelets.
\subsection{A case of coronavirus signal}
We consider in this work a strain of coronavirus associated with SARS, from a sample originally recorded in Hanoi, Vietnam since 2002-2003, See \cite{VanDerWerf}. Recall that the coronavirus is not indeed new except that it appears each time in a new form or a new state. It is for example enveloped and includes, on its surface, peplomeric structures called spicules. It may and precisely always includes proteins of unknown encoded function. Such proteins have several categories. Some are, for example, membrane glycoproteins in the form of spicules emerging from the surface of the viral envelope. They are responsible for attaching the virus to receptors in the host cell and for inducing fusion of the viral envelope with the cell membrane. Other proteins of even small variable sizes are transmembrane proteins. They play a crucial role in the budding process of coronaviruses which occurs at the level of the intermediate compartment in the endoplasmic reticulum and the Golgi apparatus. Membrane proteins constitute more than the quarter of proteins in currently sequenced complete genomes. They have a very important role in cellular processes such as the transportation of molecules and the communication between cells. Moreover, they are directly and strongly related to drugs.  More than the half of such proteins are targeted by a drug each one. Inside the membrane, the transmembrane segments may take the form of an alpha helix or the beta strand form. Generally, the size of the TM segments is of the order of 15 to 30 amino acids with a very large hydrophobic region.

When infecting a host cell, the reading frame of the viral genome is translated into a polyprotein which is cleaved by viral proteases and then releases several non-structural proteins such as RNA polymerase and ATPase helicase. These two proteins are involved in the replication of the viral genome as well as in the generation of transcripts which are used in the synthesis of viral proteins.

With the help of proteins, the virus migrates through the Golgi complex and leaves the cell and thus attaches to external bodies causing hard damages. Indeed, coronaviruses are responsible for 15 to 30\% of colds for humans and respiratory or digestive infections for animals by inducing antibodies.

The coronavirus appeared in several forms such as SARS which spread to different countries in 2002-2003. Very recently a new type of the same category of epidemics appeared originally in Hanoi, China and presents until now a challenge for humanity. The severity of these diseases is the rate or the growth of mortality in the first place and the auto-internal change of the virus althoug its external form appears the same. Determining the causative agent of the new category is now the challenge for all of humanity. More informations and ideas on such type of viruses may be found in \cite{Anand,Bonnin,Desjardins,Li,Liu,McBride,Talbot,Xu}\,.

The purpose of this work is to develop a wavelet analysis of an isolated or purified strain of human coronavirus associated with SARS already recorded and studied in \cite{VanDerWerf}. 

Recall that proteins' sequences are biological series similar and also related to DNA as they are characters' series and which also may be generated from DNA ones. The question of why preferring proteins and not DNA as others do is already discussed in \cite{Zemni1}\,. One main cause is due to the fact that proteins' sequences are more volatile. On the other hand, sequences of DNA are always issued from proteins' ones as for the example applied here. Moreover, the communication between living cells such as virus ones are always done by the intermediary of membrane and precisely transmembrane proteins. The regions of anomalies and communication constitute some type of helices which correspond to the singular and optimum points in the numerical series issued from the biological ones. See \cite{Arfaoui,Fischer-Baudoux-Woutchers,Masharifetal,Zemni1} for more details. 

In this experimental part, a reconstruction process is developed to localize the transmembrane helices of the strain of the SARS-associated coronavirus based on the hydrophobic character of the amino acids developed in \cite{KyteJ}\,. This permitted to convert proteins into time (numerical) series allowing their processing using mathematical tools to be possible (See \cite{BenMabrouk1,BenMabrouk2,Masharifetal}). The numerical conversions due to Kyte-Dolittle in \cite{KyteJ} are resumed in Table \ref{Kyte-Doolittle}. 
{\linespread{1}
	\begin{table}[h!]
		\begin{center}
			\begin{tabular}{|c|c|c|}
				\hline
				\small{Amino Acid} & \small{Kyte-Doolittle Scale} & \small{Category} \\
				\hline
				\small{Isoleucine : Ile(I)} & +4.5 & \small{Hydrophobic} \\
				\hline
				\small{Valine : Val(V)} & +4.2 & \small{Hydrophobic} \\
				\hline
				\small{Leucine : Leu(L)} & +3.8 & \small{Hydrophobic} \\
				\hline
				\small{Phenylalanine : Phe(F)} & +2.8 & \small{Hydrophobic} \\
				\hline
				\small{Cysteine : CySH(C)} & +2.5 & \small{Hydrophobic} \\
				\hline
				\small{Methionine : Met(M)}& +1.9 & \small{Hydrophobic} \\
				\hline
				\small{Alanine : Ala(A)} & +1.8 & \small{Hydrophobic} \\
				\hline
				\small{Glycine : Gly(G)} & -0.4 & \small{Neutral} \\
				\hline
				\small{Threonine : Thr(T)} & -0.7 & \small{Neutral} \\
				\hline
				\small{Serine : Ser(S)} & -0.8 & \small{Neutral}\\
				\hline
				\small{Tryptophan : Try(W)} & -0.9 & \small{Neutral} \\
				\hline
				\small{Tyrosine : Tyr(Y)} & -1.3 & \small{Neutral} \\
				\hline
				\small{Proline : Pro(P)} & -1.6 & \small{Neutral} \\
				\hline
				\small{Histidine : His(H)} & -3.2 & \small{Hydrophilic} \\
				\hline
				\small{Glutamine : Gln(Q)} & -3.5 & \small{Hydrophilic} \\
				\hline
				\small{Asparagine : Asn(N)} & -3.5 & \small{Hydrophilic }\\
				\hline
				\small{Glutamic Acid : Glu(E)} & -3.5 & \small{Hydrophilic} \\
				\hline
				\small{Aspartic Acid : Asp(D)} & -3.5 & \small{Hydrophilic} \\
				\hline
				\small{Lysine : Lys(K)} & -3.9 & \small{Hydrophilic} \\
				\hline
				\small{Arginine : Arg(R)} & -4.0 & \small{Hydrophilic}\\
				\hline
			\end{tabular}
			\caption{\small{Hydrophobicity scale of Kyte-Doolittle.}}\label{Kyte-Doolittle}
		\end{center}
\end{table}}\\
The protein strain is provided in Appendix \ref{Appendix-Coronavirus}. To illustrate the closeness of the reconstructed signal to the original one, we computed as usual the NAQE. We get the estimates provided in Table \ref{comparisontable5}.
{\linespread{1.5}
	\begin{table}[h!]
		\begin{center}
			\begin{tabular}{c|c}
				\hline\hline
				$J$&$NAQE_J$\\
				\hline
				1&$3.5108$ $10^{-12}$\\
				2&$4.3631$ $10^{-12}$\\
				3&$4.9646$ $10^{-12}$\\
				4&$5.2140$  $10^{-12}$\\
				5&$5.380$ $10^{-12}$\\
				6&$4.3631$ $10^{-12}$\\
				\hline\hline
			\end{tabular}
			\caption{NAQE estimates for the coronavirus signal using Haar-Schauder multiwavelet.}
			\label{comparisontable5}
		\end{center}
\end{table}}

\begin{figure}[h]
	\begin{center}
		\includegraphics[scale=0.95]{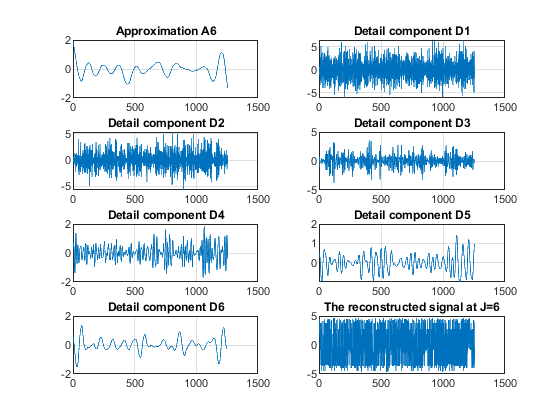}
		\caption{The decomposition of the numerized coronavirus proteins' series reconstructed with Haar-Schauder multiwavelet at the level $J=6$.}\label{Figadn}
	\end{center}
\end{figure}

Table \ref{comparisontable5} shows an optimal reconstruction reached at the level $J=6$. Next, Figure \ref{Figadn} illustrates graphically the decomposition of the numerized coronavirus proteins' series at the level $J=6$ using Haar-Schauder multiwavelet. This shows in some part the efficiency of usig multiwavelets instead of single wavelets.

\begin{center}
	\begin{figure}[h]
		\begin{center}
			\includegraphics[scale=0.5]{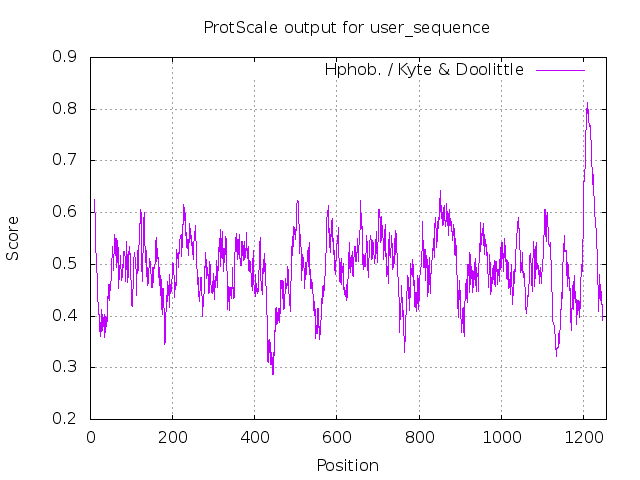}
			\caption{Kyte-Doolittle hydropathy signal for the coronavirus series}\label{KyteDolittleCorona}
		\end{center}
	\end{figure}
\end{center}

Next, as it is now well known that wavelets and multiwavelets are powerful tools to detect the transmembrane segments in proteins' series (\cite{Arfaoui,BenMabrouk1,BenMabrouk2,Masharifetal,Zemni1}) and in order to prove the applicability and thus the useful aspect of our multiwavelets we proposed to focus on the possible detection and/or prediction of alpha-helices in the considered protein. We subsequently propose to predict the locations of these regions by statistical processing applying the Haar-Schauder multiwavelet. The optima with scores greater than 1.8 (horizontal line in Figure \ref{KyteDolittleCorona}) indicate possible transmembrane regions. The window position values shown on the $x$-axis of the graph reflect the average hydropathy of the entire window, with the corresponding amino acid as the middle element of that window. Eight helices (local maxima) appear clearly.

\begin{figure}[h!]
	\centering
	\includegraphics[scale=0.95]{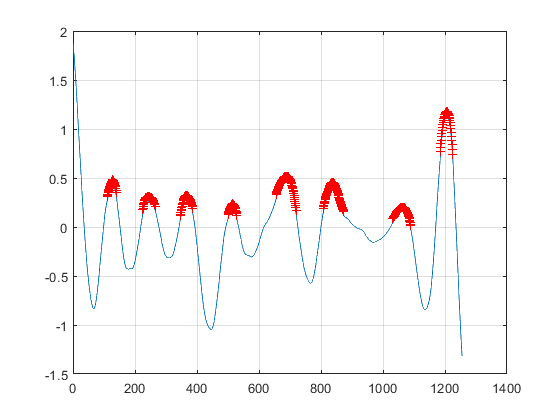}
	\caption{TMHs prediction using Haar-Schauder multiwavelet for the coronavirus signal.}\label{HaarHydrophPlot}
\end{figure}

To show the efficiency of the present method, we apply next the new explicit Haar-Schauder multiwavelet filtering method at the optimal level $J=6$. Table \ref{summarybestresults} illustrates the findings and shows 8 segments. Next, we illustrated graphically such prediction in Figure \ref{HaarHydrophPlot} which illustrates the predicted results due to the 'new' Haar-Schauder multiwavelet at the level $J=6$. It shows also 8 localized transmmembrane helices.

{\renewcommand{\arraystretch}{1}
	\begin{table}[h!]
		\begin{center}
			\begin{tabular}{c|c}
				\hline\hline
				TMHs & Haar-Schauder multiwavelet localized segments \\
				\hline
				1 & 120--134 \\
				%\hline
				2 & 233--253 \\
				%\hline
				3 & 359--373 \\
				%\hline
				4 & 505--523 \\
				%\hline
				5 & 678--699 \\
				%\hline
				6 & 824--842 \\
				%\hline
				7 & 1056--1069 \\
				%\hline
				8 & 1199--1212 \\
				\hline\hline
			\end{tabular}
			\caption{The TMHs Segments for Haar-Schauder filtering of the coronavirus signal.}\label{summarybestresults}
		\end{center}
\end{table}}
Notice that the example studied here is an important case that may be considered as a model to be applied to the new case of the coronavirus COVID-19 when a database is available which is not the case for us. We also mention that the wavelet/multiwavelet theory has proven effective in discovering and identifying abnormalities and specialfacts in biological strings such as helices, knots, .... Thus, with no laboratory study available on the chain used here and its equivalents in the new COVID-19, we intend that the current study may be applied to identify such abnormalities and other characteristics for the new virus COVID-19 chains as well as other cases.
\section{Conclusion}
In this paper, multiwavelet procedure has been developed extending the well known wavelet algorithms applied in image and signal analysis. By improving the existing ideas on multiwavelets, we constructed new ones and proved that multi-filters may be associated and applied in signal analysis with more efficient results compared to the classical ones. Error estimates as well as fast algorithms have been proved and applied on ECG signals and a coronavirus case.
\section{Appendix: The Coronavirus proteins' series strain}\label{Appendix-Coronavirus}
As we have hardy obtained the strain protein in hand we provided it in this appendix manually to be in the disposition of readers and researchers.
\begin{center}
	MPIPLLPLTLTSGSDLDRCTTPDDVQAPNYTQHTSSMRGVYYPDEIPRSDT\\
	LYLTQDLPLPPYSNVTGPHTINHTPGNPVIPPKDGIYPAATEKSNVVRGW\\
	VPGSTMNNKSQSVIIINNSTNVVIRACNPELCDNPPPAVSKPMGTQTHTMI\\
	PDNAPNCTPEYISDAPSLDVSEKSGNPKHLREPVPKNKDGPLYVYKGYQP\\
	IDVVRDLPSGPNTLKPIPKLPLGINITNPRAILTAPSPAQDIWGTSAAAYPVG\\
	YLKPTTPMLKYDENGTITDAVDCSQNPLAELKCSVKSPEIDKGIYQTSNPR\\
	VVPSGDVVRPPNITNLCPPGEVPNATKPPSVYAWERKKISNCVADYSVLYN\\
	STPPSTPKCYGVSATKLNDLCPSNVYADSPVVKGDDVRQIAPGQTGVIADY\\
	NYKLPDDPMGCVLAWNTRNIDATSTGNYNYKYRYLRHGKLRPPERDISNV\\
	PPSPDGKPCTPPALNCYWPLNDYGPYTTTGIGYQPYRVVVLSPELLNAPAT\\
	VCGPKLSTDLIKDQCVNPNPNGLTGTGVLTPSSKRPQPPQQPGRDVSDPTD\\
	SVRDPKTSEILDISPCSPGGVSVITPGTNASSEVAVLYQDVNCTDVSTAIHADQ\\
	LTPAWRIYSTGNNVPQTQAGCLIGAEHVDTSYECDIPIGAGICASYHTVSLLR\\
	STSQKSIVAYTMSLGADSSIAYSNNTIAIPTNPSISITTEVMPVSMAKTSVDCNM\\
	YICGDSTECANLLLQYGSPCTQLNRALSGIAAEQDRNTREVPAQVKQMYKTP\\
	TLKYPGGPNPSQILPPDPLKPTKRSPIEDLLPNKVTLADAGPMKQYECLGDIN\\
	ARDLICAQKPNGLTVLPPLLTDDMIAAYTAALVSGTATAGWTPGAGAALQIPP\\
	AMQMAYRPNGIGVTQNVLYENQKQIANQPNKAISQIQESLTTTSTALGKLQDV\\
	VNQNAQALNTLVKQLSSNPGAISSVLNDILSRLDKVEAEVQIDRLITGRLQSLQT\\
	YVTQQLIRAAEIRASANLAATKMSECVLGQSKRVDPCGKGYHLMSPPQAAPHG\\
	VVPLHVTYVPSQERNPTTAPAICHEGKAYPPREGVPVPNGTSWPIThQRNPPS\\
	PQIITTDNTPVSGNCDVVIGIINNTVYDPLQPELDSPKEELDKYPKNHTSPDVDL\\
	GDISGINASVVNIQKEIDRLNEVAKNLNESLIDLQELGKYEQYIKWPWYVWLGP\\
	IAGLIAIVMVTILLCCMTSCCSCLKGACSCGSCCKPDEDDSEPVLKGVKLHYT.
\end{center}
\section{Compliance with Ethical Standards}
The authors declare that no funds or grants have been received from any direction. The authors declare that no con
ict of interest for the present work. The authors declare that no animals were involved in the study and that this article does not contain any studies with human participants performed by any of the authors.
\section*{Acknowledgment}
The authors would like to recall that thee present work in its original version has been devoted to the theoretical framework of the multiwavelets processing. An application part of the work has been the object of a conference talk \cite{Zemni2}. Some extending results based on the present paper have also the object of another paper \cite{Zemni1}, although based on the present one, but has been published before it due to the long process in the publication of the present one.

\end{document}